\documentclass[journal]{IEEEtran}
\usepackage{color}
\usepackage{threeparttable}
\usepackage{srcltx}
\usepackage{psfrag}
\usepackage{subfigure}
\usepackage{epsfig}
\usepackage{amsmath}

\begin{document}
\title{Exact Finite-Difference Time-Domain Modelling of Broadband Huygens' Metasurfaces \\with Lorentz Dispersions}
\author{Tom~J.~Smy and~Shulabh~Gupta
\thanks{T. J. Smy and S. Gupta are with the Department
of Electronics, Carleton University, Ottawa, Ontario, Canada e-mail: shulabh.gupta@carleton.ca}
}

\markboth{Transactions on Antenna and Propagation 2016}%
{Shell \MakeLowercase{\textit{et al.}}: Bare Demo of IEEEtran.cls for IEEE Journals}

\maketitle

\begin{abstract}
An explicit time-domain finite-difference technique for modelling zero-thickness Huygens' metasurfaces based on Generalized Sheet Transition Conditions (GSTCs), is proposed and demonstrated using full-wave simulations. The Huygens' metasurface is modelled using electric and magnetic surface susceptibilities, which are found to follow a double-Lorentz dispersion profile. To solve zero-thickness Huygens' metasurface problems for general broadband excitations, the double-Lorentz dispersion profile is combined with GSTCs, leading to a set of first-order differential fields equations in time-domain. Identifying the exact equivalence between Huygens' metasurfaces and coupled RLC oscillator circuits, the field equations are then subsequently solved using standard circuit modelling techniques based on a finite-difference formulation. Several examples including generalized refraction are shown to illustrate the temporal evolution of scattered fields from the Huygens' metasurface under plane-wave normal incidence, in both harmonic steady-state and broadband regimes. In particular, due to its inherent time-domain formulation, a significant strength of the methodology is its ability to model time-varying metasurfaces, which is demonstrated with a simple example of a pumped metasurface leading to new frequency generation and wave amplification. 
\end{abstract}



\section{Introduction}

Metasurfaces are two dimensional arrays of sub-wavelength electromagnetic (EM) scatterers, which are the dimensional reduction of more general volumetric metamaterial structures  \cite{Metasurface_Review}\cite{meta3} and functional extensions of Frequency Selective Surfaces (FSSs)\cite{Munk_FSS}. By engineering the electromagnetic properties of the scattering particles, the metasurface can be used to manipulate and engineer the spatial wavefronts of the incident waves. In this way, they provide a powerful tool to transform incident fields into specified transmitted and reflected fields \cite{Metasurface_Synthesis_Caloz}\cite{BBParticlesTratyakov}\cite{Grbic_Metasurfaces}. More specifically, metasurfaces can either impart amplitude transformations, phase transformations or both, making them applicable in a diverse range of applications involving lensing, imaging \cite{meta2}\cite{meta3}, field transformations \cite{MetaFieldTransformation}, cloaking \cite{MetaCloak} and holograming \cite{MetaHolo}, to name a few. 

An important class of metasurfaces that has recently gathered a lot of attention is that of a \emph{Huygens' metasurface}. Huygen's metasurfaces are constructed using a 2D array of electrically small Huygen's sources, which provide perfect impedance matching to free-space, due to an optimal interaction of the electric and magnetic dipole moments ($\mathbf{p}$ and $\mathbf{m}$) constituting the metasurface \cite{Kerker_Scattering}\cite{Equalized_E_M_Tretyakov}. An efficient implementation of Huygens' metasurfaces is using all-dielectric resonator arrays \cite{Kivshar_Alldielectric}\cite{Elliptical_DMS}\cite{Grbic_Metasurfaces}, and a comprehensive review is provided in \cite{AllDieelctricMTMS}. They are ideal for providing wave-shaping, particularly in transmission, due to their non-zero reflection response. The typical applications for such Huygens' metasurfaces are, generalized refraction \cite{GeneralizedRefraction}, lensing \cite{West_DMS_Lens} and polarization control \cite{Elliptical_DMS}, to name a few. 

Metasurfaces represent abrupt phase discontinuities in space and are typically modelled as zero-thickness surfaces. This is achieved by introducing the generalized sheet transition conditions (GSTCs), accounting for exact space-discontinuities \cite{IdemenDiscont}\cite{meta2}, into frequency-domain Maxwell's equations, which can then be used to analytically synthesize the metasurface for a specified field transformation, in terms of the surface susceptibility functions \cite{Metasurface_Synthesis_Caloz}. On the other hand, an equally important problem is the computation of scattered fields from a metasurface, for its given susceptibility functions. A numerical approach has been recently presented in  \cite{CalozFDTD}, where the GSTC conditions are incorporated in the finite-difference formulation in frequency domain to accurately analyze the transmitted and reflected fields of a general zero-thickness metasurface. 

While the past works have dominantly focussed on frequency-domain solutions, there is a great interest to model and simulate broadband metasurfaces in the time-domain, where several recent works on space-time modulations  for frequency generation \cite{STGradMetasurface}\cite{ShaltoutSTMetasurface} and dispersion engineering \cite{Gupta_SpatialPhaser}\cite{Kivshar_Alldielectric}, have appeared exploring the time-domain responses of metasurfaces. With this motivation and background, a finite-difference technique is developed in this work for the specific class of a Huygens' metasurface, to compute its \emph{time-domain response} assuming a physical Lorentzian dispersion response. The combination of the time-domain GSTCs and the Lorentz response of a Huygens' metasurface, results in an explicit finite-difference formulation which can be used to simulate the transmitted and the reflected fields from the metasurface for an arbitrary time-domain excitation. Moreover, the technique can easily handle time-varying (pumped) metasurfaces, making them an ideal tool to analyze space-time modulated metasurfaces. The specific contributions of this work are: 1) Exact time-domain formulation based on GSTCs of a zero-thickness Huygens metasurface. 2) Rigorous modelling of typical sub-wavelength Huygens' metasurface using double-Lorentz dispersion model. 3) Explicit finite-difference formulation for computation of scattered fields for time-domain sources, compatible with arbitrary spatial profiles of incident waves and non-uniform Lorentz metasurfaces. 4) Determination and visualization of the scattered fields on Yee-cell grid in both transmission and reflection region. 5) Equivalence of the Huygens' metasurface with coupled RLC circuit model, useful for circuital analysis of such metasurfaces.

The paper is structured as follows. Sec. II provides the mathematical framework to analyze a Huygens' metasurface and establishes its Lorentzian dispersion response based on a typical all-dielectric unit cell model, in both the frequency and time domain. Sec. III formulates the finite-difference time-domain description of a general Lorentz Huygens' metasurface, and established its equivalence with an RLC circuit model. Sec. IV presents several demonstration examples, taking the generalized refracting metasurface as the representative case, which is followed by an example of a sinusoidally pumped metasurface. Sec. V discusses various features and limitations of the proposed approach and finally, conclusions are provided in Sec. VI.

\section{Lorentz Huygens' Metasurface}

\begin{figure}[htbp]
\begin{center}
\psfrag{A}[c][c][0.8]{$H_0$}
\psfrag{B}[r][c][0.8]{$H_r$}
\psfrag{D}[c][c][0.8]{$E_0$}
\psfrag{F}[c][c][0.8]{$E_t$}
\psfrag{E}[c][c][0.8]{$H_t$}
\psfrag{G}[c][c][0.8]{$E_r$}
\psfrag{x}[c][c][0.8]{$x$}
\psfrag{z}[c][c][0.8]{$z$}
\psfrag{J}[c][c][0.8]{$\delta = 0$}
\psfrag{a}[c][c][0.8]{$\mathbf{m_x}~[\chi_{mm}]$}
\psfrag{b}[c][c][0.8]{$\mathbf{p_y}~[\chi_{ee}]$}
\psfrag{c}[c][c][0.8]{$\mathbf{k_z}$}
\includegraphics[width=0.8\columnwidth]{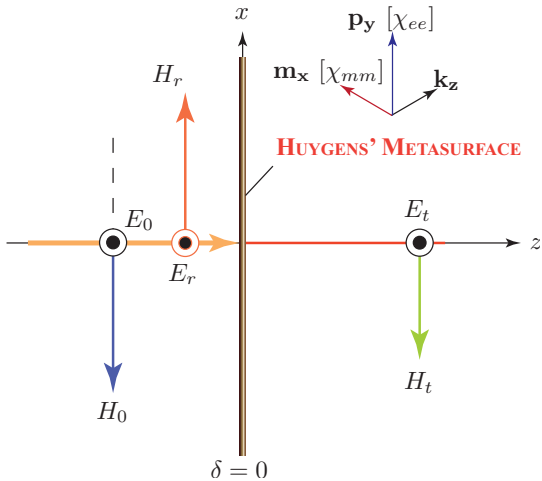}
\caption{Typical configuration of a zero-thickness Huygen's metasurface, consisting of orthogonal electric ($\mathbf{p}$) and magnetic ($\mathbf{m}$) dipole moments, excited with a normally incident plane-wave resulting in reflected and transmitted fields governed by \eqref{Eq:GSTC}.}\label{Fig:Problem}
\end{center}
\end{figure}

Consider the problem illustrated in Fig.~\ref{Fig:Problem}, where a plane wave at a frequency $\omega$ is normally incident on a metasurface. The metasurface interacts with the input field and generates both transmitted and reflected fields, in general. The metasurface is typically characterized using effective surface susceptibility densities, $\chi$, which in its most general form is a full tensor \cite{Metasurface_Synthesis_Caloz}, capturing all possible wave interactions.

A specific configuration of the metasurface consisting of a balanced and orthogonally co-located electric and magnetic dipole moments, $\mathbf{p}$ and $\mathbf{m}$, is commonly known as a \emph{Huygens' metasurface}, and is considered here throughout the paper. The electric and magnetic dipole moments are modelled using the corresponding electric and magnetic surface susceptibilities, $\chi_\text{ee}$ and $\chi_\text{mm}$. When $\chi_\text{ee}(\omega)= \chi_\text{mm}(\omega)$, the metasurface is perfectly matched to free-space. However, in the general case, both reflected and transmitted fields exist.

\subsection{Generalized Sheet Transition Conditions (GSTCs)}

A zero thickness metasurface, such as the one in Fig.~\ref{Fig:Problem}, is a space-discontinuity. The rigorous modelling of such discontinuities based on Generalized Sheet Transition Conditions (GSTCs) were developed by Idemen in \cite{IdemenDiscont}, which were later applied to metasurface problems in \cite{KuesterGSTC}. The modified Maxwell-Faraday and Maxwell-Ampere equations  can be written in the time-domain as,
\begin{subequations}\label{Eq:GSTC}
\begin{equation}
\hat{z}\times\Delta \mathbf{H} = \frac{d\mathbf{P}_{||}}{dt}
\end{equation}
\begin{equation}
\Delta \mathbf{E}\times \hat{z} = \mu_0\frac{d\mathbf{M}_{||}}{dt},
\end{equation}
\end{subequations}
\noindent where $\Delta \psi$ represents the differences between the fields on the two sides of the metasurface for all the vector component of the field $\psi$, which could be either the $\mathbf{H}$ or $\mathbf{E}$ fields. The other terms $\mathbf{P}_{||}$ and $\mathbf{M}_{||}$ represent the electric and magnetic surface susceptibility densities, \emph{in the plane of the metasurface}, which depend on the total average fields at the metasurface. A more general description and discussion can be found in \cite{Metasurface_Synthesis_Caloz}, for interested readers.

With this brief background, consider a Huygens' metasurface illuminated with a normally incident plane-wave, as shown in Fig.~\ref{Fig:Problem}. For simplicity, but without loss of generality, the problem is assumed to be 2D, where all $y-$variations are assumed to be zero. For normal incidence, the input plane-wave is given by
\begin{equation}
\mathbf{E_0}(x,z,t) = \bar{E}_0 e^{j(\omega t - kz)}  \hat{y} = E_0e^{-jkz},
\end{equation}
\noindent and $\mathbf{H_0}(x,z,t) = \mathbf{H_0}(x,z,t) /\eta_0$, where $\eta_0$ is the free-space impedance. Substituting the field quantities, assuming monochromatic excitation,
\begin{align}\label{Eq:FieldQuantities}
\Delta\mathbf{E} &= (E_t - E_r - E_0)\hat{y}\notag\\
\Delta\mathbf{H} &= (-H_t - H_r + H_0)\hat{x}\notag\\
\mathbf{P} &= \frac{\epsilon_0\chi_{ee}}{2}\frac{E_t + E_r + E_0}{2}\hat{y}\notag\\
\mathbf{M} &= \frac{\chi_{mm}}{2}\frac{-H_t + H_r -H_0}{2}\hat{x}
\end{align}
\noindent in \eqref{Eq:GSTC}, leads to 
\begin{subequations}\label{Eq:TDGSTC}
\begin{equation}
(-H_t - H_r + H_0) = \frac{\epsilon_0}{2}\frac{d}{dt}\chi_{ee}(E_t + E_r + E_0)
\end{equation}
\begin{equation}
(E_t - E_r - E_0) = \frac{\mu_0}{2}\frac{d}{dt}\chi_{mm}(-H_t + H_r -H_0).
\end{equation}
\end{subequations} 
\noindent First, let us assume that surface susceptibilities are time-independent. Under this assumption, taking the Fourier transform of the above equation and solving for the two susceptibilities, we get \cite{Metasurface_Synthesis_Caloz}
\begin{subequations}\label{Eq:Chi2TR}
\begin{equation}
\chi_{ee}(\omega) = \frac{2j}{k} \frac{\tilde{E}_t + \tilde{E}_r - \tilde{E}_0}{\tilde{E}_t + \tilde{E}_r + \tilde{E}_0} = \frac{2j}{k} \left(\frac{T + R - 1}{T + R+ 1} \right)
\end{equation}
\begin{equation}
\chi_{mm}(\omega) = \frac{2j}{k} \frac{\tilde{E}_t - \tilde{E}_r - \tilde{E}_0}{\tilde{E}_t - \tilde{E}_r + \tilde{E}_0}= \frac{2j}{k} \left(\frac{T - R- 1}{T - R + 1}\right),
\end{equation}
\end{subequations}
\noindent where $T= \tilde{E}_t/\tilde{E}_0$ and $R= \tilde{E}_r/\tilde{E}_0$ are the frequency-domain transmission and reflection transfer functions, respectively and $\tilde{\psi}$ is the Fourier transform of the time temporal fields $\psi$. Therefore, for a given transmission and reflection response of a Huygens' metasurface, \eqref{Eq:Chi2TR} can be used to compute the corresponding electric and magnetic susceptibility functions of the metasurface.

A particular case of interest is when $R=0$, i.e. reflectionless metasurface. Imposing this zero reflection condition in \eqref{Eq:Chi2TR}, leads to the condition $\tilde{\chi}_\text{ee} = \tilde{\chi}_\text{mm}$. Consequently, using either of the two sub-equations of \eqref{Eq:Chi2TR}, we get
\begin{equation}
T(\omega) = \left(\frac{2 - jk\tilde{\chi}}{2 + jk\tilde{\chi}}\right),\label{Eq:TAllPass}
\end{equation}
\noindent which is the expected all-pass transmission response of the Huygens' metasurface with $|T(\omega)|=1$, i.e. a perfect phase plate.

\subsection{Huygens' Metasurfaces}

\begin{figure}[htbp]
\begin{center}
\psfrag{a}[c][c][0.7]{$\mathbf{p}$}
\psfrag{b}[c][c][0.7]{$\mathbf{m}$}
\psfrag{c}[c][c][0.7]{$2r_1$}
\psfrag{d}[c][c][0.7]{$\Lambda$}
\psfrag{e}[c][c][0.7]{$2r_2$}
\psfrag{f}[c][c][0.7]{$t$}
\psfrag{g}[c][c][0.7]{$n_1$}
\psfrag{h}[c][c][0.7]{$n_2$}
\psfrag{x}[c][c][0.7]{$x$}
\psfrag{y}[c][c][0.7]{$y$}
\psfrag{z}[c][c][0.7]{$z$}
\psfrag{j}[c][c][0.7]{$|R| = 0$}
\psfrag{A}[c][c][0.7]{$|T|$, $|R|$~(dB)}
\psfrag{B}[c][c][0.7]{frequency~(THz)}
\psfrag{C}[c][c][0.7]{Re\{$\tilde{\chi}_\text{ee}$\}, Im\{$\tilde{\chi}_\text{ee}$\}}
\psfrag{D}[c][c][0.7]{Transmission phase $\angle T(\omega)$}
\psfrag{E}[c][c][0.7]{$|T(\omega)|$}
\psfrag{F}[c][c][0.7]{$|R(\omega)|$}
\psfrag{G}[l][c][0.6]{FEM-HFSS}
\psfrag{H}[l][c][0.6]{Lorentz \eqref{Eq:DualLZ}}
\includegraphics[width=0.95\columnwidth]{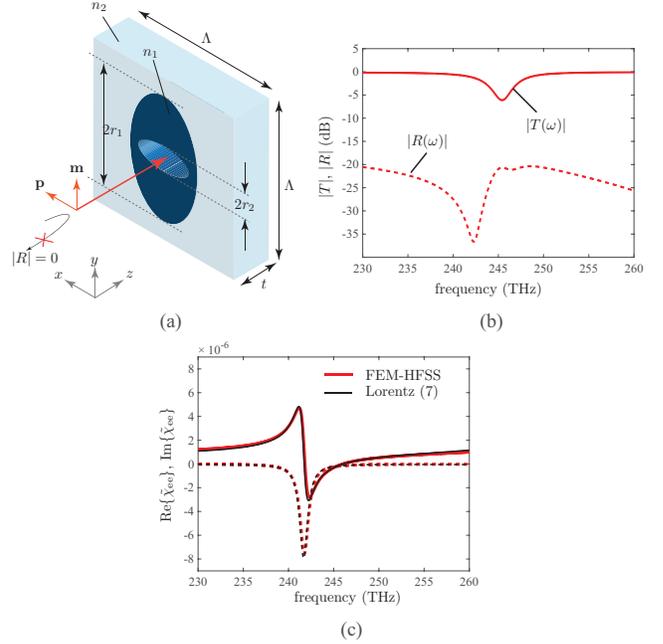}
\caption{Huygens' source metasurface based on all-dielectric resonators and its constitutive susceptibility parameters. a) Unit cell model with periodicity in both $x-$ and $y-$ directions. b) Typical amplitude transmission and reflection response, computed using FEM-HFSS, when the two dipole moments $\mathbf{p}$ and $\mathbf{m}$ are frequency aligned. c) Extracted electric susceptibility function (magnetic susceptibility not shown for simplicity) from (b) using \eqref{Eq:Chi2TR} and the Lorentz curve-fit using \eqref{Eq:DualLZ}. Design parameters: $r_1=300$~nm, $t=220$~nm, $\Lambda = 666$~nm, $n_2= 1.66$~(Silica) and $n_1=3.45$ and $\tan\delta = 0.001$~(Silicon).}\label{Fig:HuyDiMs}
\end{center}
\end{figure}

Let us next consider a typical all-dielectric metasurface unit cell, as shown in Fig.~\ref{Fig:HuyDiMs}(a), consisting of a high-dielectric cylindrical resonator (index $n_1$) embedded in a host medium of index $n_2$. The unit cell is periodic along $x-$ and $y-$directions and is subwavelength in dimension, i.e. $\Lambda< \lambda_0$ to ensure only zeroth order propagation mode. Its typical transmission and reflection amplitude response is shown in Fig.~\ref{Fig:HuyDiMs}(b), computed using FEM-HFSS. Using the computed complex transmission and reflection response of this unit cell, its corresponding surface susceptibilities can be extracted using \eqref{Eq:Chi2TR}, and are shown in Fig.~\ref{Fig:HuyDiMs}(c), where only the electric susceptibilities are shown for simplicity. Since the reflection from the metasurface is non-zero,  $\tilde{\chi}_\text{ee} \ne \tilde{\chi}_\text{mm}$ in general, across the entire frequency band of interest.

A dielectric resonator such as the one in Fig.~\ref{Fig:HuyDiMs}(a) can be modelled using a standard physical Lorentz oscillator \cite{Saleh_Teich_FP}. Following this model, it can be easily verified that the susceptibility functions $\tilde{\chi}_\text{ee}$ (and $\tilde{\chi}_\text{mm}$) of Fig.~\ref{Fig:HuyDiMs}(c), can be accurately described using a \emph{double-Lorentz function} given by
\begin{subequations}\label{Eq:DualLZ}
\begin{equation}
	\tilde{\chi}_\text{ee}(\omega) = \sum_{i=1}^2\frac{\omega_{ep,i}^2}{(\omega_{e0,i}^2 - \omega^2) + i\alpha_{e,i}\omega}  
\end{equation}
\begin{equation}
	\tilde{\chi}_\text{mm}(\omega) = \sum_{i=1}^2\frac{\omega_{mp,i}^2}{(\omega_{m0,i}^2 - \omega^2) + i\alpha_{m,i}\omega}, 
\end{equation}
\end{subequations}
\noindent as shown in Fig.~\ref{Fig:HuyDiMs}(c) for electric susceptibility, where $\omega_p$, $\omega_0$ and $\alpha$ are the plasma frequency, resonant frequency and the loss-factor of the oscillator, respectively, and subscript $e$ and $m$ denote electric and magnetic quantities. It should be noted that, while a double-Lorentz model was found sufficient to describe a typical all-dielectric unit cell like the one considered here, more complex unit cells may require more than two Lorentzian contributions.

\subsection{Lorentzian Material Response and Field Equations}

Having established the applicability of the double-Lorentzian function to describe a typical metasurface unit cell, the physical non-zero thickness unit cell of Fig.~\ref{Fig:HuyDiMs}, can now be modelled using an ideal zero-thickness Huygens' metasurface exhibiting a double-Lorentzian dispersion profile. To maintain simpler forthcoming expressions, let us first consider a single Lorentz contribution only.

The electric susceptibility, $\tilde{\chi}_\text{ee}$, in physical terms, relates the electric polarization response of the metasurface $\tilde{P}(\omega)$ produced due to the input excitation fields $\tilde{E}(\omega)$. Similarly, the magnetic susceptibility, $\tilde{\chi}_\text{mm}$, relates the magnetic polarization response of the metasurface $\tilde{M}(\omega)$ produced due to the input excitation fields $\tilde{H}(\omega)$. Mathematically, this can be expressed as
\begin{subequations}
\begin{equation}
\tilde{P}(\omega) =\tilde{\chi}_\text{ee}(\omega)\tilde{E}(\omega)=  \frac{\omega_{ep}^2}{(\omega_{e0}^2 - \omega^2) + i\alpha_e\omega} \tilde{E}(\omega)
\end{equation}
\begin{equation}
\tilde{M}(\omega) =\tilde{\chi}_\text{mm}(\omega)\tilde{H}(\omega)= \frac{\omega_{mp}^2}{(\omega_{m0}^2 - \omega^2) + i\alpha_m\omega} \tilde{H}(\omega),
\end{equation}
\end{subequations}
\noindent where a Lorentz frequency response is introduced for the susceptibilities. This can be equivalently expressed in the time-domain by inverse Fourier transforming the above equations, leading to
\begin{subequations}\label{Eq:LZ-TD}
\begin{equation}
\frac{d^2 P(t)}{dt^2} + \alpha_e \frac{dP(t)}{dt} + \omega_{e0}^2 P(t) = \omega_{ep}^2 E(t)
\end{equation}
\begin{equation}
 \frac{d^2 M(t)}{dt^2} + \alpha_m \frac{dM(t)}{dt} + \omega_{m0}^2 M(t) = - \omega_{mp}^2 H(t).
\end{equation}
\end{subequations}

To incorporate these Lorentzian responses in the time-domain GSTCs \eqref{Eq:GSTC}, \eqref{Eq:FieldQuantities} cannot be used for broadband excitations, since the polarizations can no more be expressed using simple products of surface susceptibilities and the various corresponding fields, $E$ and $H$, in the time-domain. To overcome this issue, let us introduce three time-domain polarization functions $P_t$, $P_r$ and $P_0$ corresponding to transmitted, reflected and incident fields, respectively, and use them to construct the average polarizabilities, so that
\begin{align}\label{Eq:PM_TD}
\mathbf{P}_{||} &= \epsilon_0\frac{P_t + P_r + P_0}{2}\hat{y}\notag\\
\mathbf{M}_{||} &= \frac{-M_t + M_r -M_0}{2}\hat{x}.
\end{align}
\noindent Following \eqref{Eq:LZ-TD}, each of these polarizabilities, $P_t$, $P_r$ and $P_0$, are in turn related to their corresponding fields though Lorentzian parameters, given by
\begin{subequations}\label{Eq:Set1}
\begin{equation}
\frac{d^2 P_i}{dt^2} + \alpha_e \frac{dP_i}{dt} + \omega_{e0}^2 P_i = \omega_{ep}^2 E_i(t)
\end{equation}
\begin{equation}
 \frac{d^2 M_i}{dt^2} + \alpha_m \frac{dM_i}{dt} + \omega_{m0}^2 M_i = - \frac{\omega_{mp}^2}{\eta_0} E_i(t),
\end{equation}
\end{subequations}
\noindent where the subscript $i = \text{t}\;, \text{r},\;\text{0}$, corresponds to quantities related to transmission, refection and incident waves, respectively. Next, substituting \eqref{Eq:PM_TD} into \eqref{Eq:GSTC}, and assuming plane-wave field solutions, we get  
\begin{subequations}\label{Eq:Set2}
\begin{equation}
  \mu_0\frac{dM_t}{dt}   -  \mu_0\frac{dM_r}{dt} + \mu_0\frac{dM_0}{dt} = 2(E_0 +E_r - E_t)
\end{equation}
\begin{equation}
 \frac{dP_t}{dt} +  \epsilon_0\frac{dP_r}{dt}  +  \epsilon_0\frac{dP_0}{dt} = 2c(E_0-E_t - E_r),
\end{equation}
\end{subequations}
\noindent where $c = 1/\sqrt{\mu_0\epsilon_0}$ is the speed of light in vacuum.

Finally, \eqref{Eq:Set1} and \eqref{Eq:Set2}, represent a total of eight field equations to be solved for two primary unknowns,  $E_t$, $E_r$, and five auxiliary unknowns, $P_t$, $P_r$, $P_0$, $M_t$, $M_r$, $M_0$, for a given input excitation field $E_0$. This completes the mathematical formulation of the problem.

\section{Finite Difference Formulation}

\subsection{Assembly of the 1st Order System Equations}

Let us consider the second order differential equation \eqref{Eq:LZ-TD} again, for Lorentz oscillators describing the electric and magnetic response of the metasurface. For computational simplicity, these second order differential equations can be converted into two first-differential equations, by introducing two auxiliary variables $\bar{P}$ and $\bar{M}$, defined as 
\begin{subequations}\label{Eq:Diff2Diff1}
\begin{equation}
 \bar{P} = \frac{1}{\omega_{e0}}\frac{dP}{dt} + \frac{\alpha_e}{\omega_{e0}} P
\end{equation}
\begin{equation}
\bar{M} = \frac{1}{\omega_{m0}}\frac{dM}{dt} +\frac{\alpha_m}{\omega_{m0}}M.
\end{equation}
\end{subequations}
\noindent  The rationale for using such specific definition will become clear in the next sub-section, when the equivalence of the Huygen's metasurface with a coupled circuit (RLC) model will be established. Next, using \eqref{Eq:Diff2Diff1} in \eqref{Eq:LZ-TD}, results in 
\begin{subequations}\label{Eq:Diff2Diff2}
\begin{equation}
\frac{1}{\omega_{e0}} \frac{d\bar{P}}{dt}  = \frac{\omega_{ep}^2}{\omega_{e0}^2} E - P
\end{equation}
\begin{equation}
\frac{1}{\omega_{m0}}\frac{d\bar{M}}{dt}  =-\frac{\omega_{mp}^2}{\eta_0\omega_{m0}^2} E - M
\end{equation}
\end{subequations}
\noindent where plane-wave conditions are assumed on both sides of the metasurface regions so that $H = E/\eta_0$. The general Eqs.~\eqref{Eq:Diff2Diff1} and \eqref{Eq:Diff2Diff2}, can now be written in a compact matrix form for the incident field so that
\begin{align}
	&\overbrace{\left[
	 \begin{array}{cccc}
		   \frac{1}{\omega_{e0}} & 0 & 0 & 0\\
		   0 & \frac{1}{\omega_{e0}} & 0 &0 \\
		   0& 0& \frac{1}{\omega_{m0}} & 0 \\
		  0 &0 &  0 &\frac{1}{\omega_{m0}}  \\
	\end{array} 
	\right]}^{\mathbf{W_1}^{4 \times 4}}
	\left[
	 \begin{array}{c}
		 P_0'\\
		 \bar{P}_0' \\
		 M_0'\\
		 \bar{M}_0'
	\end{array} 
	\right] + \notag\\
	&
	\overbrace{\left[
	 \begin{array}{cccc}
		\frac{\alpha_e}{\omega_{e0}} & -1 & 0 & 0\\
		 1 &  0 & 0 & 0\\
		   0& 0&\frac{\alpha_m}{\omega_{m0}} & -1\\
		  0 & 0& 1 & 0\\
	\end{array} 
	\right]}^{\mathbf{W_2}^{4 \times 4}}
	\left[
	 \begin{array}{c}
		 P_0\\
		 \bar{P}_0 \\
		 M_0\\
		 \bar{M}_0
	\end{array} 
	\right]
	=\overbrace{\left[
	 \begin{array}{c}
		 0\\
		 \frac{\omega_{ep}^2}{\omega_{e0}^2}E_0\\
		 0\\
		 -\frac{\omega_{mp}^2}{\eta_0\omega_{m0}^2}E_0
	\end{array} 
	\right]}^{\mathbf{E_1}},\label{Eq:M0}
\end{align}
\noindent where the known excitation fields $E_0$ terms are kept on the right-hand side of the matrix equation. Similarly, Eqs.~\eqref{Eq:Diff2Diff1} and \eqref{Eq:Diff2Diff2} are written in matrix form for both transmitted and reflect fields as
\begin{align}\label{Eq:Mtr}
	&\overbrace{\left[
	 \begin{array}{ccccc}
		   0 &\frac{1}{\omega_{e0}} & 0 & 0 & 0\\
		   0 &0 & \frac{1}{\omega_{e0}} & 0 &0 \\
		   0 &0& 0& \frac{1}{\omega_{m0}} & 0 \\
		  0 &0 &0 &  0 &\frac{1}{\omega_{m0}}  \\
	\end{array}
	\right]}^{\mathbf{T_1}^{4 \times 5}}
	\left[
	 \begin{array}{c}
		 E_{t,r}'\\
		 P_{t,r}'\\
		 \bar{P}_{t,r}' \\
		 M_{t,r}'\\
		 \bar{M}_{t,r}'
	\end{array} 
	\right]\notag + \\
	&
	\overbrace{\left[
	 \begin{array}{ccccc}
		0 &\frac{\alpha_e}{\omega_{e0}} & -1 & 0 & 0\\
		- \frac{\omega_{ep}^2}{\omega_{e0}^2} & 1 &  0 & 0 & 0\\
		 0&  0& 0&\frac{\alpha_m}{\omega_{m0}} & -1\\
		   \frac{\omega_{mp}^2}{\eta_0\omega_{m0}^2} & 0 & 0& 1 & 0\\
	\end{array} 
	\right]}^{\mathbf{T_2}^{4 \times 5}}
	\left[
	 \begin{array}{c}
		 E_{t,r}\\
		 P_{t,r}\\
		 \bar{P}_{t,r} \\
		 M_{t,r}\\
		 \bar{M}_{t,r}
	\end{array} 
	\right]
	=\left[
	 \begin{array}{c}
		 0\\
		 0\\
		 0\\
		 0	
	\end{array} 
	\right].
\end{align}

Next, the two equations from \eqref{Eq:Set2}, can be expressed in terms of newly introduced auxiliary variables, transforming them into a pair of linear equations, given by
\begin{subequations}
\begin{equation}
  \mu_0 \bar{M}_t   -  \mu_0\bar{M}_r  + \mu_0\bar{M}_0 + 2E_t -2E_r = 2E_0 
\end{equation}
\begin{equation}
\bar{P}_t  +  \epsilon_0\bar{P}_r +  \epsilon_0\bar{P}_0 + 2c E_t + 2cE_r   = 2c E_0,
\end{equation}
\end{subequations}
\noindent which can then be written in a compact matrix form as
\begin{align}\label{Eq:GSTC2Eq}
	\left[
	 \begin{array}{ccc}
	\mathbf{A_1} & \mathbf{A_2} &  \mathbf{A_3} \\
		\mathbf{B_1} & \mathbf{B_2} &  \mathbf{B_3}
	\end{array} 
	\right]_{2 \times 14  } [\mathbf{V}]_{14 \times 1}
		= \overbrace{\left[
	 \begin{array}{c}
	 2E_0\\
	 2cE_0\\
	\end{array} 
	\right]}^{\mathbf{E_2}},
\end{align}
\noindent where $\mathbf{V}$ is a vector consisting of all the primary and auxiliary unknown variables, i.e. 
\begin{align}
[ \mathbf{V}] = [ P_0,\;\bar{P}_0,\;M_0,\;\bar{M}_0,\;&E_t,\;P_t,\;\bar{P}_t,\;M_t,\;\bar{M}_t,\ldots \notag\\
&\;E_r,\;P_r,\;\bar{P}_r,\;M_r,\;\bar{M}_r]^\mathsf{T},
\end{align}
\noindent with $[\cdot]^\mathsf{T}$ is the matrix transpose. Finally combining \eqref{Eq:M0}, \eqref{Eq:Mtr} and \eqref{Eq:GSTC2Eq} in a single matrix, we get
\begin{align}
	&\overbrace{\left[
	 \begin{array}{ccc}
	 	\mathbf{0} & \mathbf{0} &  \mathbf{0} \\
		\mathbf{0} & \mathbf{0} &  \mathbf{0}\\
		\mathbf{W_1} & \mathbf{0} &  \mathbf{0} \\
		\mathbf{0} & \mathbf{T_1} &  \mathbf{0} \\
		 \mathbf{0}  & \mathbf{0} &  \mathbf{T_1} \\
	\end{array} 
	\right]}^{\mathbf{[C]}}
	\frac{d[\mathbf{V}]}{dt}
	+ 
	\overbrace{\left[
	 \begin{array}{ccc}
	 	\mathbf{A_1} & \mathbf{A_2} &  \mathbf{A_3} \\
		\mathbf{B_1} & \mathbf{B_2} &  \mathbf{B_3}\\
		\mathbf{W_2} & \mathbf{0} &  \mathbf{0} \\
		\mathbf{0} & \mathbf{T_2} &  \mathbf{0} \\
		 \mathbf{0}  & \mathbf{0} &  \mathbf{T_2} 
	\end{array} 
	\right]}^{\mathbf{[G]}}
	[\mathbf{V}]=\mathbf{[E]},
\end{align}
\noindent where 
\begin{align}
\mathbf{[E]} = [\mathbf{[E_2]}^\mathsf{T},\;\mathbf{[E_1]}^\mathsf{T},\; \mathbf{[0]}_{[8\times 1]}^\mathsf{T}]^\mathsf{T}.
\end{align}
\noindent The resulting $1^\text{st}$ order matrix differential equation in time, is given by
\begin{equation}
\mathbf{[C]}\frac{d\mathbf{[V]}}{dt} + \mathbf{[G]} \mathbf{[V]} = \mathbf{[E]},
\label{Eq:1stOrder}
\end{equation}
\noindent which can also be expressed equivalently in the frequency domain as
\begin{equation}
 \mathbf{[V(\omega)]} = \left(j \omega \mathbf{[C]}  + \mathbf{[G]}\right)^{-1} \mathbf{[E(\omega)]}
 \label{Eq:1stOrderFD}.
\end{equation}
\noindent This completes the necessary matrix formulation to be subsequently solved using the finite-difference technique. It should be noted that, while the formulation here is only shown for a single Lorentz oscillator for simplicity, it can be straightforwardly extended to arbitrary number of Lorentz contributions, such as the case in \eqref{Eq:DualLZ}. 

\subsection{Circuit representation}

\begin{figure}[htbp]
\begin{center}
\psfrag{A}[r][c][0.7]{$\displaystyle{\frac{\omega_{ep}^2}{\omega_{e0}^2}E_0 }$}
\psfrag{B}[c][c][0.7]{$L_p$}
\psfrag{C}[c][c][0.7]{$C_p$}
\psfrag{D}[c][c][0.7]{$R_e$}
\psfrag{E}[r][c][0.7]{$\displaystyle{\frac{- \omega_{mp}^2}{\omega_{m0}^2}\frac{E_0}{\eta_0} }$}
\psfrag{F}[c][c][0.7]{$L_m$}
\psfrag{G}[c][c][0.7]{$C_m$}
\psfrag{H}[l][c][0.7]{$R_m$}
\psfrag{I}[c][c][0.7]{$V_e$}
\psfrag{J}[l][c][0.7]{$V_m$}
\includegraphics[width=0.9\columnwidth]{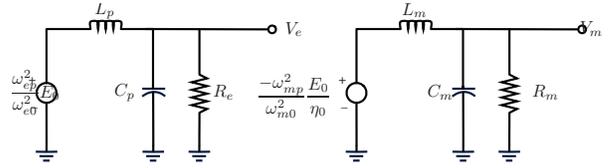}
\caption{A coupled RLC resonator circuit model as an equivalent circuit model representation of the Huygens' metasurface of Fig.~\ref{Fig:Ckt}.}\label{Fig:Ckt}
\end{center}
\end{figure}

It is not a coincidence that the set of equations described by \eqref{Eq:1stOrder} is in a standard form used widely for circuit analysis. Both the field equations \eqref{Eq:Set2} and the Lorentzian material behaviour \eqref{Eq:Set1} are easily represented by circuits.  Specifically, the polarization responses \eqref{Eq:Set1} can be simply seen as a set of coupled RLC oscillators. For example in Fig.~\ref{Fig:Ckt}, two circuits representing the response of $M$ and $P$ due to the incident field $E_0$ are shown for which the corresponding circuit equations in matrix form are given by
\begin{align}
	& \left[
	 \begin{array}{cccc}
		   C_p & 0 & 0 & 0\\
		   0 & L_p & 0 &0 \\
		   0& 0& C_m & 0 \\
		  0 &0 &  0 & L_m  \\
	\end{array}
	\right]
	\left[
	 \begin{array}{c}
		 V_{p0}'\\
		 I_{p0}' \\
		 V_{m0}'\\
		 I_{m0}'
	\end{array} 
	\right]
	+ \notag\\
	&\left[
	 \begin{array}{cccc}
		\frac{1}{R_e} & -1 & 0 & 0\\
		 1 &  0 & 0 & 0\\
		 0& 0& \frac{1}{R_m} & -1\\
		 0 & 0& 1 & 0\\
	\end{array}
	\right]
	\left[
	 \begin{array}{c}
		 V_{p0}\\
		 I_{p0} \\
		 V_{m0}\\
		 I_{m0}
	\end{array} 
	\right]
	=\left[
	 \begin{array}{c}
		 0\\
		 \frac{\omega_{ep}^2}{\omega_{e0}^2}\\
		 0\\
		 \frac{\omega_{ep}^2}{\omega_{e0}^2} \frac{1}{\eta_0}
	\end{array} 
	\right] E_0,
\end{align}
\noindent where the unknowns are the currents through the inductors and the output voltages. If we identify $V_{p0} \leftrightarrow P_0$, $V_{m0} \leftrightarrow M_0$, $I_{p0} \leftrightarrow \bar P_0$ and $I_{m0} \leftrightarrow \eta_0 \bar M_0$ and set $L = C = 1/\omega_0$ and $R = \frac{\omega_0}{\alpha}$, we obtain the matrices derived previously, for example, \eqref{Eq:M0}. The other polarizations ($P_t$, $P_r$, $M_t$ and $M_r$) would have identical circuit implementations. The field matrix equations are also easily implemented using circuits and the form of \eqref{Eq:1stOrder} is then obtained by using standard circuit modelling techniques. This equation can now be solved in either the frequency domain (if not time varying) or it can be solved using a finite-difference formulation using standard techniques for the general case.  

\subsection{Extension for Time Varying Metasurface Parameters} 

The analysis above implicitly assumes static surface parameters. However, with a simple extension, an active surface can be easily modelled. Actively pumping the surface would modify the parameters of the Lorentzians relating $M$ and $P$ to the $E$ fields. To investigate this, we decided to modify the resonant frequency of the resonators in time. To achieve this, the RLC circuit description was used and the capacitance $C$ was varied as $C(t) = C_0 + A \sin(\omega_p t)$. To include this in the formulation described above is straightforward. The static capacitor is replaced with a non-linear capacitor described by the equations: $Q_c = C(t) V_c $ and $I_c = dQ/dt$, where $Q_c$ is the charge stored in the capacitor, $V_c$ is the voltage across the capacitor and $I_c$ the current through the capacitor. This, consequently, requires a minor modification to the circuit equations, as shown in \eqref{Eq:LCVaryin}.
\begin{figure*}[!t]
\normalsize
\begin{equation}\label{Eq:LCVaryin}
 \left[
	 \begin{array}{cccccc}
		   0 & 0 & 1 & 0 & 0 & 0\\
		   0 & L_p & 0 &0 &0 & 0 \\
		   0 & 0 & 0 &0 &0& 0\\
		   0& 0& 0 & 0 & 0 & 1\\
		  0 &0 &  0 & L_m &0 &0\\
		  0 & 0 & 0 &0 &0& 0\\
	\end{array}
	\right]
	\left[
	 \begin{array}{c}
		 V_{p0}'\\
		 I_{p0}' \\
		 Q_{p0}' \\
		 V_{m0}'\\
		 I_{m0}'\\
		 Q_{m0}' \\
	\end{array} 
	\right]
	+ \left[
	 \begin{array}{cccccc}
		\frac{1}{R_e} & -1 & 0 & 0&0&0\\
		 1 &  0 & 0 & 0 & 0 & 0\\
		 0 &  0 & 1 & 0 & 0 & 0\\
		 0& 0& 0 &\frac{1}{R_m} & -1 & 0\\
		 0 & 0& 1 & 0 &0 &0\\
		 0 &  0 & 0 & 0 & 0 & 1\\
	\end{array}
	\right]
	\left[
	 \begin{array}{c}
		 V_{p0}\\
		 I_{p0} \\
		 Q_{p0}' \\
		 V_{m0} \\
		 I_{m0} \\
		 Q_{p0}' \\
	\end{array} 
	\right]
	=\left[
	 \begin{array}{c}
		 0\\
		 0\\
		 C_p(t)V_{p0}\\
		 0\\
		 0\\
		 C_m(t)V_{m0}\\
	\end{array} 
	\right]
	+ 
	\left[
	 \begin{array}{c}
		 0\\
		 \frac{\omega_{ep}^2}{\omega_{e0}^2}\\
		 0\\
		 0\\
		 \frac{\omega_{ep}^2}{\omega_{e0}^2} \frac{1}{\eta_0}\\
		 0\\
	\end{array} 
	\right] E_0.
\end{equation}
\hrulefill
\vspace*{4pt}
\end{figure*}

 This results in a introduction of a new non-linear term to the system equations leading to
\begin{equation}
\mathbf{[C]}\frac{d\mathbf{[V]}}{dt} + \mathbf{[G]} \mathbf{[V]} = \mathbf{[B(V)]} + \mathbf{[E]},
\label{Eq:MNA}
\end{equation}
\noindent Finally, writing the explicit finite-difference form of \eqref{Eq:MNA} using the Trapezoidal rule, we get
\begin{align}
\mathbf{[V]_i} & = \left(\mathbf{[C]} +  \frac{\Delta t\mathbf{[G]}}{2}\right)^{-1}\notag\\
&\left(\mathbf{[C]}\mathbf{[V]_{i-1}} - \frac{\Delta t\mathbf{[G]} \mathbf{[V]_{i-1}}}{2}\right. \notag\\
&\left.+ \Delta t \left[ \frac{\mathbf{[E]_i} +\mathbf{[E]_{i-1}}}{2} + \frac{\mathbf{[B]_i} +\mathbf{[B]_{i-1}}}{2}\right] \right)
\end{align}
\noindent where $i$ is the index denoting the current time stamp. If the surface is static $\mathbf{B(V)} = \mathbf{0}$ and for a uniform time step we can simply use an LU decomposition of $\mathbf{H} = \mathbf{C} + \Delta t/2 \mathbf{G}$ to solve the matrix equation at each time step. However, if $\mathbf{B}$ is non zero due to time modulation of the metasurface, Newton-Raphson iteration will be needed at each time step to solve the non-linear nature of the equations and $\mathbf{H}$ will need to include the Jacobian of $\mathbf{B}$. If needed, a large number of techniques have been developed to solve equations of the form of \eqref{Eq:MNA} that can be easily exploited \cite{mna,mna2,spice}. 

\section{Examples}

\subsection{Simulation Setup}

\begin{figure}[tbp]
\begin{center}
\psfrag{A}[c][c][0.8]{$E_r,\; H_r$}
\psfrag{B}[c][c][0.8]{$E_t,\; H_t$}
\psfrag{C}[c][c][0.8]{$E_0(x, z= 0_{-})$}
\psfrag{D}[c][c][0.8]{$E_t(x, z= 0_{+})$}
\psfrag{E}[c][c][0.8]{$z=0$}
\psfrag{F}[c][c][0.8]{$E_r(x, z= 0_{-})$}
\psfrag{X}[c][c][0.8]{$x$}
\psfrag{Z}[c][c][0.8]{$z$}
\psfrag{G}[l][c][0.8]{$\delta \rightarrow 0$}
\includegraphics[width=\columnwidth]{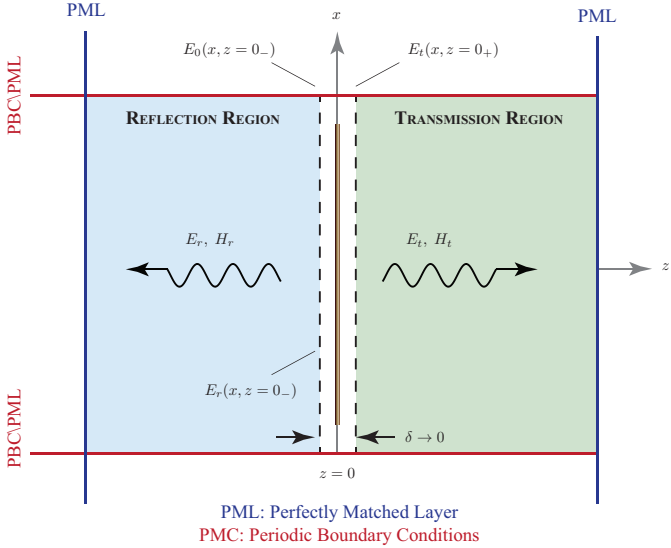}
\caption{Simulation setup for the finite-difference technique developed in Sec. III showing the numerical domain and the surrounding boundary conditions. Shaded regions on the left and right represent the 2D discretized region where Maxwell's equations are solved on a conventional Yee cell.}\label{Fig:Setup}
\end{center}
\end{figure}

To illustrate the methods described above two simulation cases will be used. The first case is a simulation of a uniform metasurface with the material response described by a Lorentzian response. For this case the problem is one dimensional and we have only to solve \eqref{Eq:1stOrder} in either the time or frequency domain. This case will be used initially to present the response of the surface to an incident pulse and then subsequently be extended to a time varying (or pumped) metasurface, which can only be solved in the time domain using \eqref{Eq:MNA}. The second case involves the simulation of a surface with a graded material response to achieve generalized refraction. The purpose of this case is to study the propagation of the transmitted and reflected fields from such a metasurface, for which analytical solutions are readily available.

The propagation from the surface was modelled as shown in Fig. \ref{Fig:Setup}. Two regions were defined in which the EM propagation was modelled using a standard 2D Yee cell model \cite{taflove2000computational}. The metasurface is placed at $z = 0$ and a region modelling the reflected wave is situated on left. The second region models the propagation of the transmitted field and is situated on the right side of the metasurface. The discretized metasurface model was used as boundary condition for the two regions with the transmitted field $H_{ty}$ being imposed on the left side of the region modelling the transmission. To model the reflection from the metasurface, $H_{ry}$ is similarly imposed on the right side of the region on the left. Further, as shown in Fig.~\ref{Fig:Setup}, the top and bottom boundaries are periodic and the two left- and right-end boundaries have perfectly matched layers (PMLs). The propagation of the incident field was not modelled and a uniform plane wave $E_0$ was imposed on the surface at $z = 0_-$. The metasurface equations were then solved using the same time steps as the propagation regions. 

\begin{figure}[tbp]
\begin{center}
\psfrag{a}[c][c][0.6]{Field Amplitude}
\psfrag{b}[c][c][0.6]{Phase (rad)}
\psfrag{c}[c][c][0.6]{frequency, $f$ (THz)}
\psfrag{d}[l][c][0.5]{$|E_t/E_0|_\text{m}$}
\psfrag{e}[l][c][0.5]{$|E_r/E_0|_\text{m}$}
\psfrag{f}[l][c][0.5]{$|E_t/E_0|_\text{mm}$}
\psfrag{g}[l][c][0.5]{$|E_r/E_0|_\text{mm}$}
\psfrag{h}[l][c][0.5]{$\angle \{E_t/E_0\}$}
\psfrag{i}[l][c][0.5]{$\angle \{E_r/E_0\}$}
\includegraphics[width=\columnwidth]{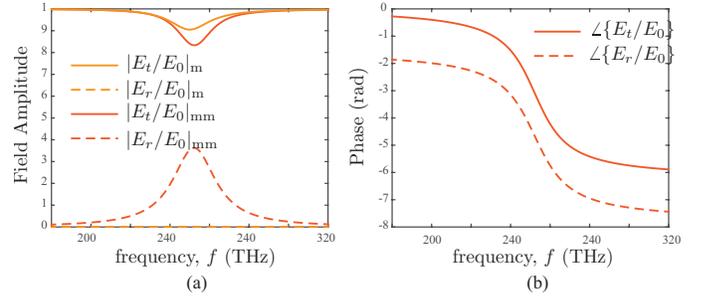}
\caption{Frequency response of a typical uniform Lorentz Huygens' metasurface for a matched (denoted with a subscript \emph{m}) and an unmatched configuration (\emph{mm}) a) Transmission and reflection response, b) transmission and reflection phase for the case of mismatched configuration.}\label{Fig:SurfFRes}
\end{center}
\end{figure}

\begin{figure*}[tbp]
\begin{center}
\psfrag{a}[c][c][0.6]{E-Field Volts/m}
\psfrag{b}[c][c][0.6]{time, $t$ (ps)}
\psfrag{c}[c][c][0.6]{$E_0$}
\psfrag{d}[c][c][0.6]{$E_t$}
\psfrag{e}[c][c][0.6]{$E_r$}
\includegraphics[width=1.75\columnwidth]{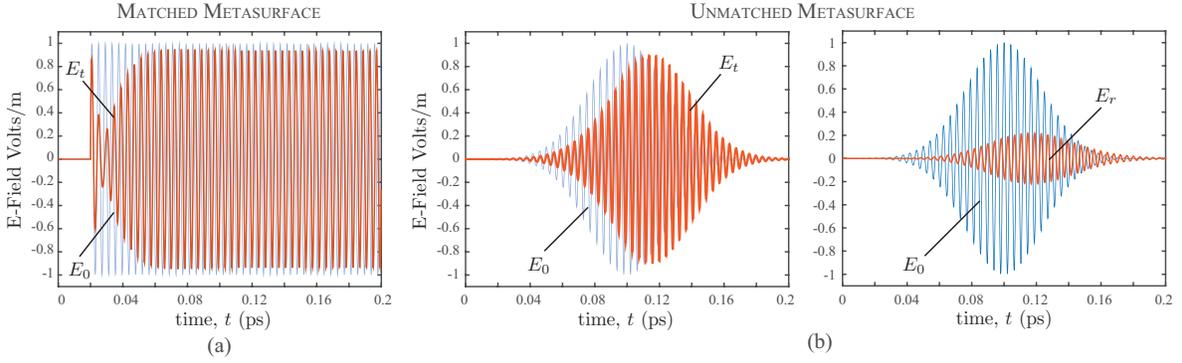}
\caption{Time-domain response of a Lorentz Huygens' metasurface. a) Transmitted field from a matched metasurface excited with stepped input with a modulation frequency of $f_0 = 240$~THz. b) Transmitted and reflected fields from a mismatched metasurface under Gaussian pulse excitation. The incident Gaussian pulse has a unit magnitude and a width of 33.3~fs. }\label{Fig:PulseProp}
\end{center}
\end{figure*}

Values for the material response were extracted from FEM-HFSS simulations of the silicon structure described in Fig. \ref{Fig:HuyDiMs}, taken as a representative test case here. As shown before, two Lorentzian contributions are need to accurately model the response. Unless otherwise noted the parameters used for the Lorentzian material responses in the following are:  For first Lorentzian contribution, $f_{e0} = f_{m0} = 250$~THz, $f_{wp} = f_{mp} = 48$~GHz and $\alpha_e = \alpha_m = 7.54\times10^{12}$. And for second the Lorentzian contribution, $f_{e0} = f_{m0} = 350$~THz, $f_{wp} = f_{mp} = 183$~GHz and $\alpha_e = \alpha_m = 7.54\times10^{12}$.

\subsection{Pulse Propagation through the Metasurface}

For the case of a uniform metasurface, \eqref{Eq:1stOrderFD} can be used to characterize the metasurface as a function of frequency. Fig.~\ref{Fig:SurfFRes} presents the magnitude and phase response of the metasurface for two cases. The first case is that of matched surface where both resonators are identical, i.e. $\tilde{\chi}_{ee}(\omega) = \tilde{\chi}_{mm}(\omega)$. It can be seen that the reflection is zero as expected with some loss near the resonant frequency of the resonators. The response is not symmetrical due to the presence of the $2^\text{nd}$ Lorentzian at 350~THz. The second case shows the result of mismatching the $P$ and $M$ responses by offsetting the resonant frequency of $M$ by 5~THz. As can be seen, now a substantial reflection is present. The corresponding phase response is also shown for both cases in Fig. \ref{Fig:SurfFRes} where it should be noted that both the transmitted and reflected fields undergo almost a full 2$\pi$ phase shift as frequency $f$ is swept from 180~THz to 320~THz. 

The response of this uniform metasurface to an incident pulse is shown in Fig.~\ref{Fig:PulseProp}, computed using the proposed finite-difference formulation of \eqref{Eq:1stOrder}. Again two cases are shown. Fig.~\ref{Fig:PulseProp}(a) shows the response of the metasurface to a step input of a modulated field (250~THz). Initially the surface is transparent as the resonators are ``at rest'' and there is full transmission of the field. There is then a short period as the resonators absorb energy and come to a harmonic steady-state. At this point the surface is transmitting with a small loss. At no point in the transient response is there any reflection, as the resonators never become mismatched. The second case is shown in Fig.~\ref{Fig:PulseProp}(b) and is the response of the mismatched surface to an incident modulated Gaussian pulse ($f = 250$~THz and a pulse width of 33.3~fs). Fig.~\ref{Fig:PulseProp}(b) shows the transmitted and reflected pulses, where they are both delayed and slightly distorted due to the dispersion of the metasurface. These results were then compared and confirmed for benchmarking purposes, with the frequency domain model (or the Fourier propagation model) of the metasurface, i.e. $v_\text{out}(t) = \mathcal{F}^{-1}[\mathcal{F}\{v_\text{in}(t)\} T(\omega)]$, where $T(\omega)$ is the metasurface transfer function shown in Fig.~\ref{Fig:SurfFRes}.  

\subsection{Huygens' Metasurface for Generalized Refraction}

In this section we will use the method described above to analyze a metasurface designed to produce a refraction of an incident plane wave. The material response of the surface will be ``engineered'' by creating a distribution of resonators with varying $\omega_0$ which could be achieved by optimizing the geometry of the resonator, for instance. First, we shall look at the general response and design of a matched surface using the harmonic steady-state response and then we shall present the transient response of a surface under matched and mismatched conditions, computed using our proposed finite-difference technique using \eqref{Eq:1stOrder}. 

\subsubsection{Surface response at harmonic steady-state}

\begin{figure*}[tbp]
\begin{center}
\psfrag{A}[c][c][0.7]{Surface Susceptibility $\tilde{\chi}_{ee}(\omega)$}
\psfrag{B}[c][c][0.7]{Surface Susceptibility $\chi(x)$}
\psfrag{C}[c][c][0.7]{Frequency $f$ (THz)}
\psfrag{D}[c][c][0.7]{distance $x$~($\mu$m)}
\psfrag{E}[l][c][0.5]{$2\tan(\phi_0x)/k$}
\psfrag{F}[l][c][0.5]{$\chi_{ee}(\omega_0)$ (Lossy)}
\psfrag{G}[l][c][0.5]{$\chi_{ee}(\omega_0)$ (Lossless)}
\psfrag{H}[l][c][0.5]{Im$\{\cdot\}$}
\psfrag{J}[l][c][0.5]{Re$\{\cdot\}$}
\psfrag{K}[c][c][0.7]{$\omega_0$}
\psfrag{a}[c][c][0.6]{$\omega_0\times 10^{15}$ (rad/s)}
\psfrag{b}[c][c][0.6]{Field Amplitude}
\psfrag{c}[c][c][0.6]{Phase (rad)}
\psfrag{y}[c][c][0.6]{distance $x$~($\mu$m)}
\psfrag{d}[l][c][0.6]{$|E_t/E_0|_\text{m}$}
\psfrag{e}[l][c][0.6]{$|E_t/E_0|_\text{mm}$}
\psfrag{f}[l][c][0.6]{$|E_r/E_0|_\text{mm}$}
\psfrag{g}[l][c][0.6]{$\angle \{E_t/E_0\}^\text{m}$}
\psfrag{h}[l][c][0.6]{$\angle \{E_t/E_0\}^\text{um}$}
\psfrag{i}[l][c][0.6]{$\angle \{E_r/E_0\}^\text{um}$}
\psfrag{x}[c][c][0.8]{$x$}
\psfrag{z}[c][c][0.8]{$z$}
\psfrag{p}[l][c][0.8]{$\Delta\phi = 2\pi$}
\psfrag{q}[c][c][0.8]{$\Delta x = 10~\mu$m}
\includegraphics[width=2\columnwidth]{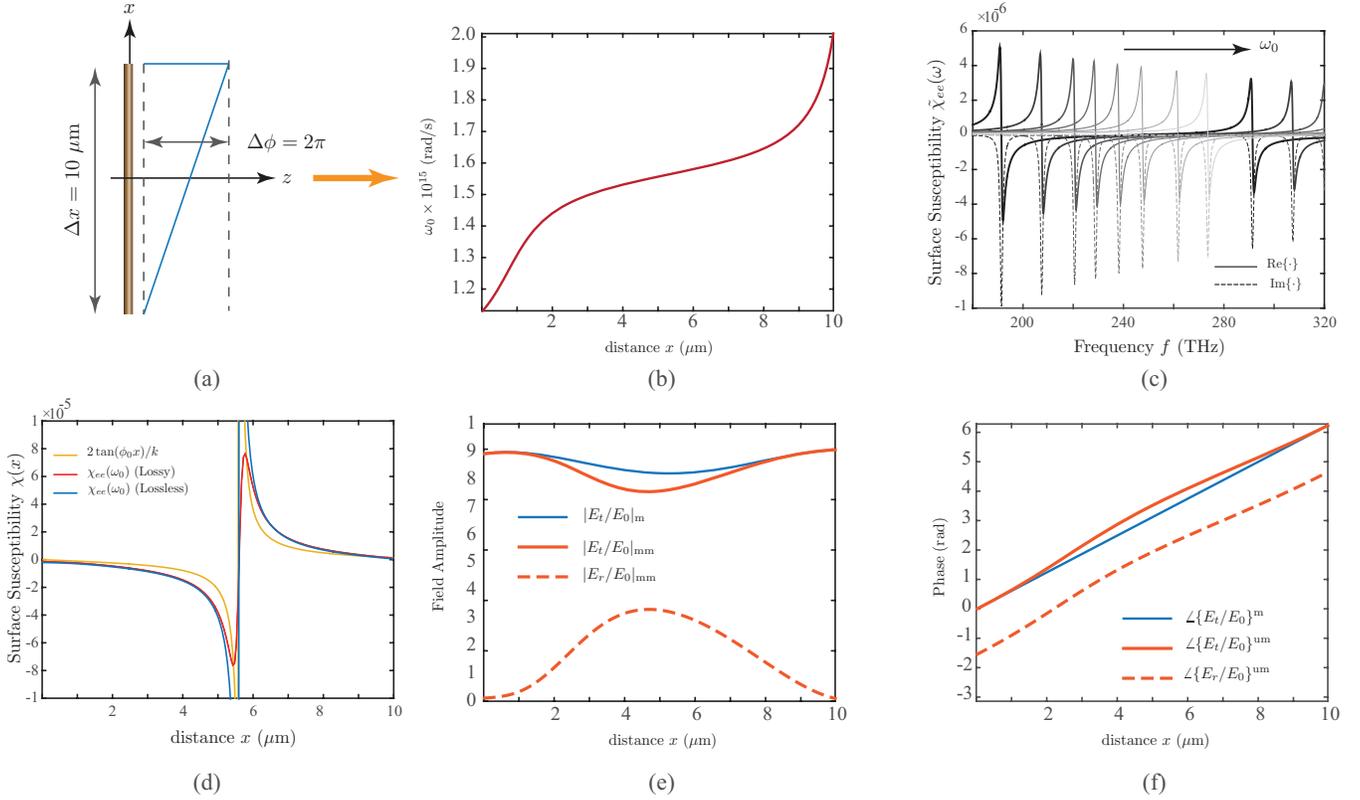}
\caption{A generalized refracting Lorentz metasurface exhibiting a linear-phase across the metasurface. a) Configuration, b) The corresponding variation of the resonant frequency $\omega_0(x)$ of the Lorentz oscillator associated with the electric dipole moment $\mathbf{p}$. c) Double-Lorentz susceptibility density function $\tilde{\chi}_{ee}$ for various resonant frequencies $\omega_0$ d) Susceptibility density function $\chi_{ee}(x)$ for a given excitation frequency $\omega$. e) and f) The corresponding transmission and reflection amplitude/phase response for the matched and mismatched configurations. }\label{Fig:SurfRes}
\end{center}
\end{figure*}

Fig.~\ref{Fig:SurfRes} shows the response of graded metasurface where the metasurface was engineered to produce a linearly increasing phase shift across the surface as illustrated in Fig.~\ref{Fig:SurfRes}(a). This effect was achieved by using the response of the metasurface shown in Fig.~\ref{Fig:SurfFRes} and determining a distribution of $\omega_0$ that will produce this linear phase response along the metasurface for a matched surface. This distribution is shown in Fig.~\ref{Fig:SurfRes}(b). In Fig.~\ref{Fig:SurfRes}(c), the corresponding susceptibilities for 10 resonators equally spaced along the surface are shown as function of frequency. Finally, by solving \eqref{Eq:1stOrderFD} at each point on the surface, we can obtain the response of the surface in harmonic steady-state. 

It is first useful to establish the metasurface response for an ideal matched case for a given excitation frequency. If we define, 
\begin{align}
	\tilde{\chi}_{ee}(x) = \tilde{\chi}_{ee}(x)= \tilde{\chi}_{ee}(x) = \tilde{\chi} = \frac{2\tan(\phi_0x)}{k}, \label{Eq:IdealGEChi}
\end{align}
\noindent it can be verified by substituting the above form of $\tilde{\chi}$ in \eqref{Eq:TAllPass} that the transmission function takes a simpler form $T(\omega) = e^{-2j\phi_0x}$, so that the output wave is given by 
\begin{equation}
\mathbf{E_t}(x,z,t)  = E_0 e^{j\omega t} e^{-j(kz + 2\phi_0x)}  \hat{y}, \label{Eq:RefrWave}
\end{equation}
\noindent which represents an output refracted-wave in the $x-z$ plane. Having noted that, Fig.~\ref{Fig:SurfRes}(d) shows the real part of the engineered $\tilde{\chi}_{ee}(x)$ for our generalized refracting metasurface, for both lossy and lossless Lorentzian oscillators, and compares them with the ideal response of \eqref{Eq:IdealGEChi}. As can be seen from this comparison, the engineered Lorentzian response is a good proxy for the ideal tangent function of \eqref{Eq:IdealGEChi}. 

Finally, the corresponding transmission and reflection transfer function distributions along the metasurface are next shown in Fig.~\ref{Fig:SurfRes}(e) and (f), for the two cases of matched and mismatched metasurfaces. For the matched case, the reflection, as expected, is zero throughout the metasurface and the transmission is close to unity with some variation in the absorption due to loss in the material. While the phase response of the matched case is perfectly linear as designed, the mismatched case (i.e. shift in $f_0$ of 5 THz between electric and magnetic resonances) exhibits a substantial and a non-uniform reflection across the metasurface with a noticeable non-linear phase response.

\subsubsection{Transient surface response and propagation}

\begin{figure}[htbp]
\begin{center}
\psfrag{a}[c][c][0.6]{distance $z$~($\mu$m)}
\psfrag{b}[c][c][0.6]{distance $x$~($\mu$m)}
\psfrag{c}[c][c][0.6]{$t_1=0.4fs$}
\psfrag{d}[c][c][0.6]{$t_2=10fs$}
\psfrag{e}[c][c][0.6]{$|E_t/E_0|_\text{mm}$}
\psfrag{f}[c][c][0.6]{$|E_y|$}
\psfrag{g}[c][c][0.6]{$|H_z|$}
\psfrag{h}[c][c][0.6]{$|H_x|$}
\psfrag{x}[c][c][0.7]{$\boxed{\chi_{ee} = \chi_{mm}}$}
\includegraphics[width=\columnwidth]{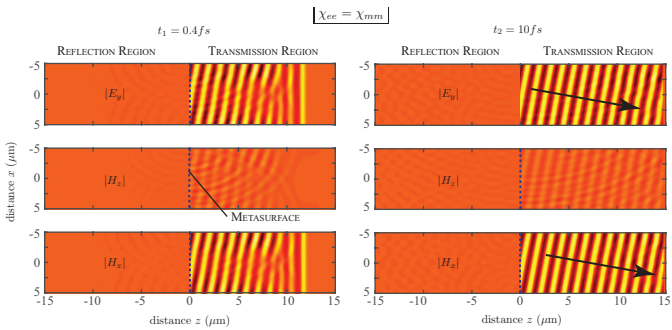}
\caption{Time-domain computed transmitted and reflected fields from generalized refracting metasurface, incident with a normally incident plane-wave, for the matched case when $\chi_{ee} = \chi_{mm}$, at two time instants $t_1 = 0.4$~fs and $10$~fs. Colorplot is linear in scale.}\label{Fig:FieldSolMatched}
\end{center}
\end{figure}

Figure~\ref{Fig:FieldSolMatched} shows the computed fields, at two time instants for a matched generalized refracting metasurface of Fig.~\ref{Fig:SurfRes} excited with a normally incident plane-wave. At $t_1 = 0.4$~fs during early excitation times, the surface is still in the initial stage of interaction with the wave and the transmitted wavefronts are not fully refracted. However, as the material response increases, the phase response of the surface begins to create the refraction or the ``tilting'' of the incident plane wave, as clearly seen in the wavefronts close to the metasurface in Fig.~\ref{Fig:FieldSolMatched}. At a later time, $t_2 =10$~fs which is long enough for the material to come to harmonic steady-state, the linear phase response of the circuit can clearly be seen manifesting itself as clean rotation of the incident plane wave at an angle $\theta = 5.99^\circ$, which matches with the refraction angle expected from the theoretical prediction from \eqref{Eq:RefrWave} using a tangent function for $\tilde{\chi}$ ($\theta_{ideal} = 6.08^\circ$). Since this is a matched surface, no reflection would be expected and that is indeed what is seen. 

%

\begin{figure}[htbp]
\begin{center}
\psfrag{a}[c][c][0.6]{distance $z$~($\mu$m)}
\psfrag{b}[c][c][0.6]{distance $x$~($\mu$m)}
\psfrag{c}[c][c][0.6]{$t_1=0.4fs$}
\psfrag{d}[c][c][0.6]{$t_2=10fs$}
\psfrag{e}[c][c][0.6]{$|E_t/E_0|_\text{mm}$}
\psfrag{f}[c][c][0.6]{$|E_y|$}
\psfrag{g}[c][c][0.6]{$|H_z|$}
\psfrag{h}[c][c][0.6]{$|H_x|$}
\psfrag{x}[c][c][0.7]{$\boxed{\chi_{ee} \ne \chi_{mm}}$}
\includegraphics[width=\columnwidth]{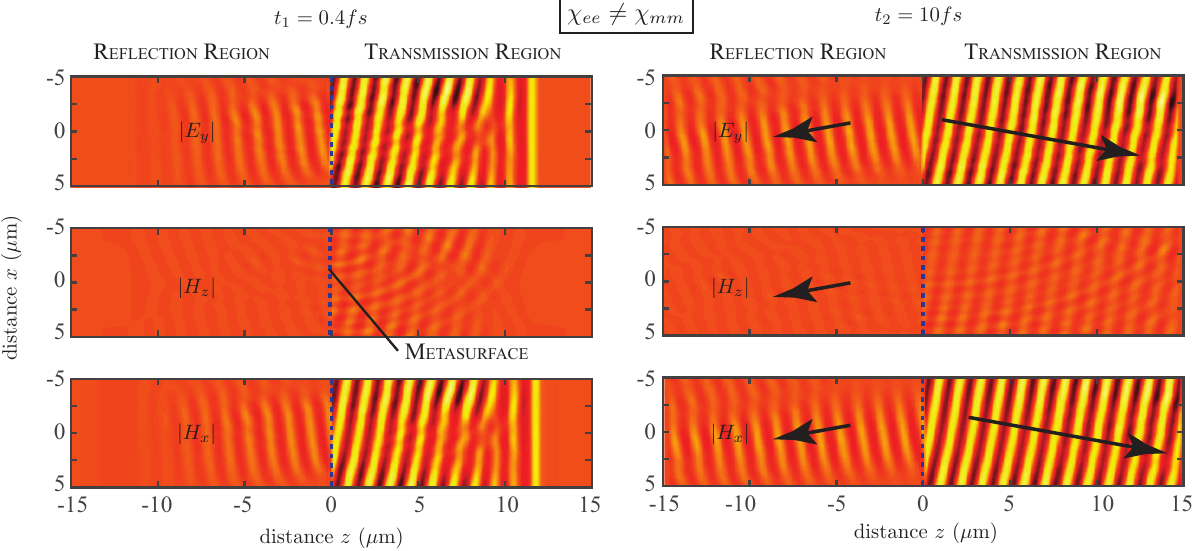}
\caption{Time-domain computed transmitted and reflected fields from generalized refracting metasurface, incident with a normally incident plane-wave, for the unmatched case when $\chi_{ee} \ne \chi_{mm}$, at two time instants $t_1 =0.4$~fs and $t_2 = 10$~fs. Colorplot is linear in scale.}\label{Fig:FieldSolUnmatched}
\end{center}
\end{figure}

Next, the second case of a mismatched metasurface is shown in Fig. \ref{Fig:FieldSolUnmatched}. As previously, computed fields at time instants at $t_1 = 0.4$~fs and $t_2 =10$~fs are shown. For the case of the short simulation time, the metasurface response is similar to that of matched case, showing an initial starting period during which the material response comes to steady-state. Of course, due to the mismatch of the $P$ and $M$ resonators, a reflected field is seen as well. As would be expected from Fig. \ref{Fig:SurfRes}(e), the reflection is spatially distributed leading to some diffraction effects. At time $t_2$, the metasurface response has come to a harmonic steady-state and the refraction of the incident field is clearly seen. However, due to the mismatch in the resonators, the transmitted and reflected waves are no longer perfect flat plane waves and exhibits noticeable amplitude variations. Finally, As expected the reflected field is also rotated by the same angle as the transmitted fields. It should be noted that in both Figs.~\ref{Fig:FieldSolMatched} and \ref{Fig:FieldSolUnmatched}, diffractive effects can clearly be seen in middle plot of $H_z$, which can be attributed to the slight discontinuity in the fields at the top and bottom of the metasurface which are linked by the periodic boundary conditions in the simulation setup. 

\subsection{Time-varying Pumped Metasurface}

The final simulation illustrates the ability of the method to model time-varying or pumped metasurfaces. In this simulation a uniform surface was modelled using \eqref{Eq:MNA} to take into account the temporal variations in surface susceptibilities of the metasurface. Specifically, the surface parameters $\chi$'s, were varied in time by using a time dependent capacitance to define the resonant frequency of the material. In addition, a mismatched surface was modelled with a 5~THz shift in the resonance frequency of the $P$ and $M$ resonances. The capacitance was varied at a pumping frequency of $f_p =280$~THz with the peak magnitude variation of 10\%.

\begin{figure}[tbp]
\begin{center}
\psfrag{a}[c][c][0.6]{Field Amplitude}
\psfrag{b}[c][c][0.6]{time (ps)}
\psfrag{c}[c][c][0.6]{$E_0$}
\psfrag{d}[c][c][0.6]{$E_t$}
\psfrag{e}[c][c][0.6]{$E_r$}
\psfrag{f}[l][c][0.5]{FFT\{$E_0$\}}
\psfrag{g}[l][c][0.5]{FFT\{$E_t$\}}
\psfrag{h}[l][c][0.5]{FFT\{$E_r$\}}
\psfrag{j}[c][c][0.6]{frequency $f$ (THz)}
\psfrag{k}[c][c][0.6]{Normalized spectrum (dB)}
\includegraphics[width=\columnwidth]{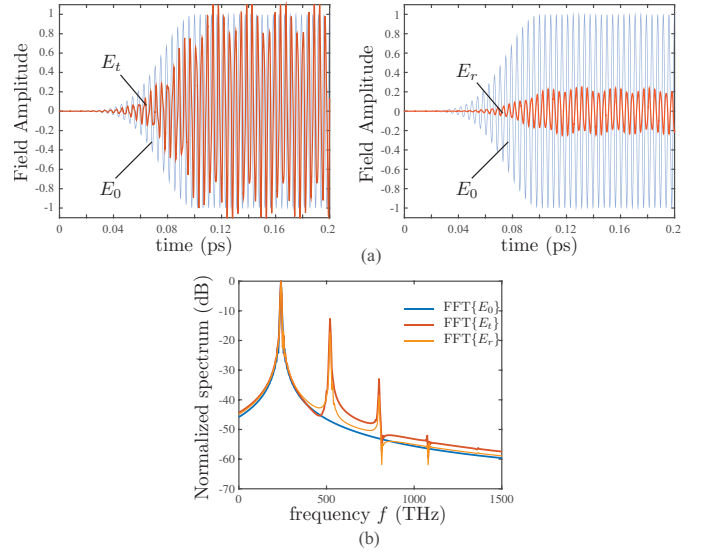}
\caption{Transmitted and reflected fields from a time-varying metasurface excited with an incident step function, where the resonant frequency of the Lorentzians characterizing the material is varied in time via $C = C_0(1 + 0.1 \sin(\omega_p t))$ where $f_p = 280$THz is the pumping frequency. The rising edge of the step was Gaussian in shape with a width of $33.3$~fs. a) Transmitted and reflected fields. b) Field spectrums computed using FFT.}\label{Fig:TimeVaryingMS}
\end{center}
\end{figure}

Figure~\ref{Fig:TimeVaryingMS} shows the transmitted and reflected fields from the metasurface excited with a modulated time-step function. The initial rise of the step was modelled using Gaussian profile with a width of 33.3~fs and the modulation frequency of 250~THz. Figure~\ref{Fig:TimeVaryingMS}(a) shows the temporal evolution of the fields in both reflection and transmission. Two effects can be noticed from these results: 1) A strong amplitude modulation in both reflected and transmitted fields, indicating the generation of new temporal frequencies; 2) An increase in transmission field amplitudes compared to the input, indicating gain in the system. Both frequency generation and wave amplification are typical effects encountered in the general class of space-time modulations system \cite{OlinerST}\cite{CassidyST}, and can be expected in this pumped metasurface example. To confirm the generation of new frequencies, the spectrum of both reflected and transmitted fields was computed using Fast-Fourier Transforms (FFTs) and is shown in Fig.~\ref{Fig:TimeVaryingMS}(b). The creation of new frequency content at multiples of $f_p = 280$~THz is clearly seen and thus confirmed.

\section{Discussions \& Future Work}

The finite-difference numerical method proposed and described above has number of key features leading to advantages over other existing approaches. Firstly, a primary need for efficient design and analysis of a metasurface is the development of simple physically motivated models that are also accurate and robust. The surface model presented herein is physically accurate, flexible and can be implemented with ease. Due to its finite-difference form, no issue of numerical instability arises. On the other hand, the alternative time and frequency domain methods primarily used in full wave EM simulators such as FEM-HFSS, FDTD-CST and FEM-COMSOL are hampered by the complexity implied in the simulation of an engineered metasurface, the need to assume a finite thickness for the surface and the inability to solve the corresponding time-varying problems. The proposed technique can moreover be easily coupled to propagation models or other system descriptions and thus is well suited as a part of computer-aided design (CAD) tool, as has also been done here by combining it with the Yee-cell model. Also to be noted that, while the examples have been taken at optical frequencies for illustration, the proposed technique is obviously applicable for modelling microwave Huygens' metasurfaces as well. Secondly, while the paper assumed relatively straitforward incident field distributions (i.e. normally incident plane-waves) throughout for simplicity, the proposed method is fully compatible with wide variety of input fields of arbitrary spatial and temporal characteristics. Since there is no implicit assumptions in the proposed method, it can easily handle spatially distributed and complex time varying fields ranging from Gaussian beams to vectorial beams \cite{VectorBeams}. Thirdly, a significant strength of the methodology is its ability to model time-varying or pumped metasurfaces. The example presented in the paper was the simple time modulation of the metasurface, however, both time and space varying phenomena, i.e. $\chi_{ee}(\mathbf{r}, t)$ and $\chi_{mm}(\mathbf{r}, t)$, could easily be modelled, making the proposed technique, ideal for analyzing wide range of problems in the general context of space-time modulated metasurfaces. Finally, the metasurface response was assumed to be  linear Lorentzian response, in this work. However, for the time-domain model, this assumption is not necessary and it could be extended to include more general non-linear metasurfaces for large incident fields where $\chi = \chi(E)$ and the linear assumption breaks down.

The technique as presented also has two major limitations, however, both of them can be simply overcome. In the examples considered herein, the metasurface itself was one dimensional and was coupled into 2D propagation regions. However, the extension to 3D problems is straight-forward with a discretized 2D surface being described by an array of resonators and the corresponding set of full-vectorial field equations. This metasurface then would be used as a boundary condition for the ful 3D EM simulation region. Such a simulation framework could then be used to model either infinite surfaces (periodic boundary conditions) or finite surfaces (absorbing boundary conditions) as also is the case in this work. 

A second limitation is that the method has assumed an incident field from one side only and then either transmission or reflection fields are computed using the GSTCs. A more general and complex problem will involve scattering from objects surrounding the metasurface in both transmission and reflection regions. Considering that the metasurface is linear with respect to the field response, the general problem of metasurface in a scattering environment, can, in principle, be overcome by simply having \emph{two} auxiliary mathematical surfaces, one for left to right field transmission and the other for right to left transmission. The reflected and transmitted fields produced on both sides of these auxiliary surfaces, would then be added to produce the total fields propagating from the surface in both directions. This extension to the method would thus enable the simulation of multiple metasurfaces, for instance, coupled by intermediate propagation regions. 

\section{Conclusions}

An explicit time-domain finite-difference technique for modelling zero-thickness Huygens' metasurfaces based on GSTCs, has been proposed and demonstrated using full-wave simulations. The Huygens' metasurface has been modelled using electric and magnetic surface susceptibilities, which were found to follow a double-Lorentz dispersion profile. To solve zero-thickness Huygens' metasurface problems for general broadband excitations, the double-Lorentz dispersion profile was integrated with GSTCs, leading to sets of first-order differential fields equations in time-domain. Identifying the exact equivalence between Huygens' metasurfaces and coupled RLC oscillator circuits, the field equations were then subsequently solved using standard circuit modelling techniques based on finite-difference formulation. Using this proposed technique, several examples including generalized refraction have been shown to illustrate the temporal evolution of scattered fields from the Huygens' metasurface under plane-wave normal incidence, in both harmonic steady-state and broadband regimes. In particular, due to its inherent time-domain formulation, a significant strength of the methodology is its ability to model time-varying metasurfaces, which has been demonstrated with a simple example of a pumped metasurface leading to new frequency generation and wave amplification. 

The proposed technique is based on simple physically motivated Lorentz models that are accurate and able to handle wide variety of incident beams and can be easily extended to solve for a full 3D metasurface problems. The proposed technique thus represents an efficient design and analysis tool for broadband metasurfaces and can be easily coupled to other propagation models and system descriptions, as a part of computer-aided design (CAD) tool.


\begin{thebibliography}{10}
\providecommand{\url}[1]{#1}
\csname url@samestyle\endcsname
\providecommand{\newblock}{\relax}
\providecommand{\bibinfo}[2]{#2}
\providecommand{\BIBentrySTDinterwordspacing}{\spaceskip=0pt\relax}
\providecommand{\BIBentryALTinterwordstretchfactor}{4}
\providecommand{\BIBentryALTinterwordspacing}{\spaceskip=\fontdimen2\font plus
\BIBentryALTinterwordstretchfactor\fontdimen3\font minus
  \fontdimen4\font\relax}
\providecommand{\BIBforeignlanguage}[2]{{%
\expandafter\ifx\csname l@#1\endcsname\relax
\typeout{** WARNING: IEEEtran.bst: No hyphenation pattern has been}%
\typeout{** loaded for the language `#1'. Using the pattern for}%
\typeout{** the default language instead.}%
\else
\language=\csname l@#1\endcsname
\fi
#2}}
\providecommand{\BIBdecl}{\relax}
\BIBdecl

\bibitem{Metasurface_Review}
C.~Holloway, E.~F. Kuester, J.~Gordon, J.~O'Hara, J.~Booth, and D.~Smith, ``An
  overview of the theory and applications of metasurfaces: The two-dimensional
  equivalents of metamaterials,'' \emph{IEEE Antennas Propag. Mag.}, vol.~54,
  no.~2, pp. 10--35, April 2012.

\bibitem{meta3}
N.~Yu and F.~Capasso, ``Flat optics with designer metasurfaces,'' \emph{Nature
  Materials}, vol.~13, April 2014.

\bibitem{Munk_FSS}
B.~A. Munk, \emph{Frequency Selective Surface: Theory and Design}.\hskip 1em
  plus 0.5em minus 0.4em\relax New York: Wiley, 2000.

\bibitem{Metasurface_Synthesis_Caloz}
K.~Achouri, M.~Salem, and C.~Caloz, ``General metasurface synthesis based on
  susceptibility tensors,'' \emph{IEEE Trans. Antennas Propag.}, vol.~63,
  no.~7, pp. 2977--2991, July 2015.

\bibitem{BBParticlesTratyakov}
V.~S. Asadchy, I.~A. Faniayeu, Y.~Ra'di, S.~A. Khakhomov, I.~V. Semchenko, and
  S.~A. Tretyakov, ``Broadband reflectionless metasheets: Frequency-selective
  transmission and perfect absorption,'' \emph{Phys. Rev. X}, vol.~5, p.
  031005, Jul 2015.

\bibitem{Grbic_Metasurfaces}
C.~Pfeiffer and A.~Grbic, ``Metamaterial huygens' surfaces: Tailoring wave
  fronts with reflectionless sheets,'' \emph{Phys. Rev. Lett.}, vol. 110, p.
  197401, May 2013.

\bibitem{meta2}
C.~Holloway, E.~F. Kuester, J.~Gordon, J.~O'Hara, J.~Booth, and D.~Smith, ``An
  overview of the theory and applications of metasurfaces: The two-dimensional
  equivalents of metamaterials,'' \emph{Antennas and Propagation Magazine,
  IEEE}, vol.~54, no.~2, pp. 10--35, April 2012.

\bibitem{MetaFieldTransformation}
S.~A. Tretyakov, ``Metasurfaces for general transformations of electromagnetic
  fields,'' \emph{Philosophical Transactions of the Royal Society of London A:
  Mathematical, Physical and Engineering Sciences}, vol. 373, no. 2049, 2015.

\bibitem{MetaCloak}
Y.~Yang, H.~Wang, Z.~X. F.~Yu, and H.~Chen, ``A metasurface carpet cloak for
  electromagnetic, acoustic and water waves,'' \emph{Scientific Reports.},
  no.~6, pp. 1--6, Jan. 2016.

\bibitem{MetaHolo}
G.~Zheng, H.~Muhlenbernd, M.~Kenney, G.~Li, T.~Zentgraf, and S.~Zhang,
  ``Metasurface holograms reaching 80\% efficiency,'' \emph{Nat. Nanotech.},
  no.~43, pp. 308--312, Feb. 2015.

\bibitem{Kerker_Scattering}
M.~Kerker, \emph{The Scattering of Light and Other Electromagnetic
  Radiation}.\hskip 1em plus 0.5em minus 0.4em\relax Academic Press, New York,
  1969.

\bibitem{Equalized_E_M_Tretyakov}
E.~Saenz, I.~Semchenko, S.~Khakhomov, K.~Guven, R.~Gonzalo, E.~Ozbay, and
  S.~Tretyakov, ``Modeling of spirals with equal dielectric, magnetic, and
  chiral susceptibilities,'' \emph{Electromagnetics}, vol.~28, no.~7, pp.
  476--493, 2008.

\bibitem{Kivshar_Alldielectric}
M.~Decker, I.~Staude, M.~Falkner, J.~Dominguez, D.~N. Neshev, I.~Brener,
  T.~Pertsch, and Y.~S. Kivshar, ``High-efficiency dielectric huygens'
  surfaces,'' \emph{Adv. Opt. Mater.}, vol.~3, no.~6, pp. 813--820, 2015.

\bibitem{Elliptical_DMS}
A.~Arbabi, Y.~Horie, M.~Bagheri, and A.~Faraon, ``Dielectric metasurfaces for
  complete control of phase and polarization with subwavelength spatial
  resolution and high transmission,'' \emph{Nat. Nanotech.}, vol.~10, p.
  937Ð943, Jul 2015.

\bibitem{AllDieelctricMTMS}
S.~Jahani and Z.~Jacob, ``All-dielectric metamaterials,'' \emph{Nature
  Nanotechnology}, vol.~2, no.~11, pp. 23--36, Jan 2016.

\bibitem{GeneralizedRefraction}
N.~Yu, P.~Genevet, M.~A. Kats, F.~Aieta, J.-P. Tetienne, F.~Capasso, and
  Z.~Gaburro, ``Light propagation with phase discontinuities: Generalized laws
  of reflection and refraction,'' \emph{Science}, vol. 334, no. 6054, pp.
  333--337, 2011.

\bibitem{West_DMS_Lens}
P.~R. West, J.~L. Stewart, A.~V. Kildishev, V.~M. Shalaev, V.~V. Shkunov,
  F.~Strohkendl, Y.~A. Zakharenkov, R.~K. Dodds, and R.~Byren, ``All-dielectric
  subwavelength metasurface focusing lens,'' \emph{Opt. Express}, vol.~22,
  no.~21, pp. 26\,212--26\,221, Oct 2014.

\bibitem{IdemenDiscont}
M.~M. Idemen, \emph{Discontinuities in the Electromagnetic Field}.\hskip 1em
  plus 0.5em minus 0.4em\relax John Wiley \& Sons, 2011.

\bibitem{CalozFDTD}
Y.~Vahabzadeh, K.~Achouri, and C.~Caloz, ``Simulation of metasurfaces in finite
  difference techniques,'' \emph{Arxiv 1602.04086v1}, Feb. 2016.

\bibitem{STGradMetasurface}
Y.~Hadad, D.~L. Sounas, and A.~Alu, ``Space-time gradient metasurfaces,''
  \emph{Phys. Rev. B}, vol.~92, p. 100304, Sep 2015.

\bibitem{ShaltoutSTMetasurface}
A.~Shaltout, A.~Kildishev, and V.~Shalaev, ``Time-varying metasurfaces and
  lorentz non-reciprocity,'' \emph{Opt. Mater. Express}, vol.~5, no.~11, pp.
  2459--2467, Nov 2015.

\bibitem{Gupta_SpatialPhaser}
S.~Gupta, K.~Achouri, and C.~Caloz, ``All-pass metasurfaces based on
  interconnected dielectric resonators as a spatial phaser for real-time analog
  signal processing,'' in \emph{2015 IEEE Conference on Antenna Measurements
  Applications (CAMA)}, Nov 2015, pp. 1--3.

\bibitem{KuesterGSTC}
E.~F. Kuester, M.~A. Mohamed, M.~Piket-May, and C.~L. Holloway, ``Averaged
  transition conditions for electromagnetic fields at a metafilm,'' \emph{IEEE
  Transactions on Antennas and Propagation}, vol.~51, no.~10, pp. 2641--2651,
  Oct 2003.

\bibitem{Saleh_Teich_FP}
B.~E.~A. Saleh and M.~C. Teich, \emph{Fundamentals of Photonics}, 2nd~ed.\hskip
  1em plus 0.5em minus 0.4em\relax Wiley-Interscience, 2007.

\bibitem{mna}
C.-W. Ho, A.~Ruehli, and P.~Brennan, ``The modified nodal approach to network
  analysis,'' \emph{IEEE Transactions onCircuits and Systems}, vol.~22, no.~6,
  pp. 504 -- 509, Jun. 1975.

\bibitem{mna2}
U.~Wali, R.~Pal, and B.~Chatterjee, ``On the modified nodal approach to network
  analysis,'' \emph{Proceedings of the IEEE}, vol.~73, no.~3, pp. 485--487,
  March 1985.

\bibitem{spice}
T.~Quarles, A.~Newton, D.~Pederson, and A.~Sangiovanni-Vincentelli, \emph{SPICE
  3 Version 3F5 User's Manual}, Dept. of EECE, Univ. of California, Berkeley.

\bibitem{taflove2000computational}
A.~Taflove and S.~Hagness, ``Computational electrodynamics: The
  finite-difference time-domain method,'' 2000.

\bibitem{OlinerST}
E.~S. Cassedy and A.~A. Oliner, ``Dispersion relations in time-space periodic
  media: Part i stable interactions,'' \emph{Proceedings of the IEEE}, vol.~51,
  no.~10, pp. 1342--1359, Oct 1963.

\bibitem{CassidyST}
E.~S. Cassedy, ``Dispersion relations in time-space periodic media part ii
  unstable interactions,'' \emph{Proceedings of the IEEE}, vol.~55, no.~7, pp.
  1154--1168, July 1967.

\bibitem{VectorBeams}
Q.~Zhan, ``Cylindrical vector beams: from mathematical concepts to
  applications,'' \emph{Adv. Opt. Photon.}, vol.~1, no.~1, pp. 1--57, Jan 2009.

\end{thebibliography}
\end{document}